\documentclass[twocolumn,pre,nofootinbib]{revtex4}

\newcommand{\beq}{\begin{eqnarray}}
\newcommand{\eeq}{\end{eqnarray}}

\usepackage{graphicx,amsmath,amssymb,tabularx}

\usepackage{color} 

\begin{document}

\title{The connection between statics and dynamics of spin glasses}
\author{Matthew Wittmann}
\author{A.~P.~Young}
\affiliation{Department of Physics, University of California, Santa Cruz, California 95064, USA}


\begin{abstract}
We present results of numerical simulations on a one-dimensional Ising spin
glass with long-range interactions. Parameters of the model are chosen such
that it is a proxy for a short-range spin glass \textit{above}
the upper critical dimension (i.e.\ in the mean-field regime).
The system is quenched to a
temperature well below the transition temperature $T_c$ and the growth of
correlations is observed. The spatial decay of the correlations at distances
less than the dynamic correlation length $\xi(t)$ agrees quantitatively with
the predictions of a \textit{static} theory, the ``metastate'', evaluated according to
the replica symmetry breaking (RSB) theory. We also compute the dynamic
exponent $z(T)$ defined by $\xi(t) \propto t^{1/z(T)}$ and find that it is
compatible with the mean-field value of the \textit{critical} dynamical exponent for
short range spin glasses.
\end{abstract}

\maketitle

Experimental measurements on a system at finite-temperature involve a
\textit{time average}. Dynamics is harder to calculate than statics, so, in
theoretical work,
the time average is usually replaced by a static calculation
using statistical mechanics in which one \textit{sums over all
configurations} with
the Boltzmann probability distribution. Most systems are ergodic, so
theory agrees with
experiment even though different averages are performed.  One situation where
more care is needed is that of a phase transition where symmetry is
spontaneously broken. A simple case is the Ising ferromagnet, which has just
two ordered states below the transition temperature $T_c$, the ``up'' spin
state with a net positive magnetization, and the 
``down'' state. 
On cooling the system will choose one of these symmetry-related states and
acquire a non-zero magnetization. However, the Boltzmann sum will
(unphysically) include both the up and down states and give zero net
magnetization.
The ``up'' and ``down'' states in
the Ising ferromagnet are called ``pure states'' in the 
literature~\cite{newman:03}.
In pure states,
correlation functions have a ``clustering'' property which means
that ``connected'' correlation vanish at infinity, i.e. 
\begin{equation}
\lim_{|{\bf r}_i - {\bf r}_j| \to \infty} \left(\, \langle S_i S_j\rangle - \langle S_i
\rangle \langle S_j \rangle \, \right) = 0 \, . 
\label{cluster}
\end{equation}
This does not occur in the
Boltzmann average since the second term
is zero, so the combination of
``up'' and ``down'' is not a pure state; rather it is a ``mixed
state''. 


However, in spin glasses, which have disorder and ``frustration'', the
situation is much more complicated. \textit{Dynamically}, below the spin glass transition
temperature $T_c$ a macroscopic system is not in thermal equilibrium because
relaxation times are much too long.
Rather, in a typical experiment
the system is quenched from a high temperature to a temperature below $T_c$
and the subsequent dynamic evolution of the system is observed.
\textit{For statics}, the state (or states) of
thermal equilibrium are very complicated and are not related to any symmetry.
As for the
ferromagnet we would like to find a static calculation which will
predict the experimental behavior, at least to some extent. In this paper we
show \textit{quantitatively} that the theoretical idea called the
``metastate''~\cite{newman:97,aizenman:90},
combined with the technique of ``replica symmetry
breaking'' (RSB)~\cite{parisi:80,parisi:83}
provides such a description for spin glasses, at least in high dimensions, where
the critical behavior is described by mean field theory.

Since the clustering property in Eq.~\eqref{cluster}
is convenient one would like to also
describe spin glasses in terms of pure states. This can be done (in principle)
by taking a very
large system, applying some boundary condition on it, and studying the
correlations in a relatively small window far from the
boundary~\cite{newman:03,read:14}.
This procedure is
repeated for many different boundary conditions.

The question of whether there are many pure states or just one (a
time-reversed pair in the absence of a magnetic field) in spin glasses has
been very controversial
~\cite{fisher:87,fisher:88,parisi:80,parisi:83},
If there are
many, one needs to do some sort of statistical average over them, which is
called a ``metastate'', for which different but equivalent formulations have
been given by Newman and Stein~\cite{newman:97} and by Aizenman and Wehr
(AW)~\cite{aizenman:90}. In the AW metastate, one considers a scale $M$,
intermediate between the window size $W$ and the system size $L$. The
metastate-averaged state (MAS) is obtained by computing correlation functions
in the window in which an average is performed not only over the spins but
also over the bonds in the ``exterior'' region between $M$ and
$L$~\cite{aizenman:90,read:14,manssen:14}.

For the infinite-range model of Sherrington and
Kirkpatrick (SK)~\cite{sherrington:75}, Parisi's exact
solution~\cite{parisi:80,parisi:83}, solved by 
RSB,
has many pure states in a sense that was clarified later by
Newman and Stein~\cite{newman:97}, see also Read~\cite{read:14}.

The critical behavior of a realistic spin
glass is expected to be the same as that of the
SK~\cite{sherrington:75} model
in dimension $d$
greater than the ``upper critical dimension'', $d_u$, which is equal to six.
However, this does not necessarily mean that the RSB description of the spin
glass phase \textit{below} $T_c$ also applies for $d > 6$
~\cite{fisher:87,fisher:88,newman:97}.
Nonetheless Read~\cite{read:14} has
computed the spatial fluctuations in a finite-dimensional model below $T_c$, assuming
mean-field (Gaussian) fluctuations, and the metastate description coming from
Parisi's~\cite{parisi:80,parisi:83}
RSB solution of the SK model. Spin correlations are
found~\cite{dedominicis:06,marinari:00a,read:14} to decay
with a power of the distance, due to the averaging over many pure states
(which are unrelated by symmetry) in the metastate, i.e.
\begin{equation}
\langle S_i S_j \rangle^2_\text{MAS} \propto 1 /
r_{ij}^{\alpha_s}\quad \text{with}\  \ 
\alpha_s = d - 4,
\label{alpha_s}
\end{equation}
where ``s'' refers to ``static,
``MAS'' stands for metastate-averaged state, and sites $i$ and $j$ are
in the window far from the boundary.
The result in Eq.~\eqref{alpha_s} had been obtained earlier in
Ref.~\cite{marinari:00a} from an RSB calculation working in the zero overlap
sector. 

We emphasize that the calculation leading to Eq.~\eqref{alpha_s} is a
\textit{static} one.  Is it possible to relate it to experiments (or numerical
simulations), which concern (non-equilibrium) \textit{dynamics}? Many
simulations~\cite{rieger:93,kisker:96,marinari:96,yoshino:02,bellettietal:09,manssen:14}
have been carried out in which a spin glass is quenched to below
$T_c$ and the resulting dynamics analyzed. It is found that fluctuations
reach a steady state on length scales smaller than a
dynamic correlation length $\xi(t)$ which is found, empirically, to grow with
a power of $t$ like
\begin{equation}
\xi(t) \propto t^{1 / z(T)} \, ,
\label{xizT}
\end{equation}
where the non-equilibrium dynamic exponent $z(T)$ is found to vary, roughly,
like $1/T$ and becomes close to the critical dynamical exponent, $z_c$, for
$T =
T_c$,~\cite{rieger:93,kisker:96,marinari:96,yoshino:02,bellettietal:09,manssen:14,katzgraber:05b,zTc}
\begin{equation}
{1 / z(T) } \simeq (T / T_c)\, z_c \, .
\label{zT}
\end{equation}

At distances less than $\xi(t)$ correlations are observed to fall off with a
power of distance leading to the following scaling prediction
\begin{equation}
C_4(r_{ij}, t) \equiv \left[\, \langle S_i(t) S_j(t) \rangle^2 \, \right]
= {1 \over r_{ij}^{\alpha_d}}\, f\left(r_{ij} \over \xi(t)\right) \, ,
\label{C4rt}
\end{equation}
where ``d'' refers to ``dynamic''. Here the thermal average squared,
$\langle \cdots \rangle^2$, is
performed by simulating two copies of the system with the same interactions,
initialized with different random spin configurations. Spin
configurations of the two copies at the same time are used in Eq.~\eqref{C4rt}.
Use of two copies
provides an unbiased estimate of this thermal average. The second average,
$[\cdots]$, is over the
bonds. We will also average over all pairs of sites a given distance $r$
apart. For $r_{ij} \ll \xi(t)$ one finds $f(x \to 0) = \text{const.}$ so
\begin{equation}
C_4(r_{ij}, t) \propto {1 / r_{ij}^{\alpha_d}}\, \qquad (r_{ij} \ll \xi(t))\, .
\label{alpha_d}
\end{equation}

Clearly, the non-equilibrium dynamics is generating a sampling of the pure states.
To our knowledge, White and Fisher~\cite{white:06} were the first to point out the
similarity of this sampling and the metastate average for statics. They use
the term ``maturation metastate'' to describe the ensemble of states generated
dynamically on scales less than $\xi(t)$ following a quench,
and ``equilibrium metastate'' for the static
metastate discussed earlier. Here we will use terms ``dynamic'' and ``static'' 
to describe these two metastates. Subsequently Manssen et
al.~\cite{manssen:15} emphasized the similarity between the two metastates and
suggested that they might actually be equivalent, in which case $\alpha_s$ in
Eq.~\eqref{alpha_s} would equal $\alpha_d$ in Eq.~\eqref{alpha_d}. The
rationale behind
this hypothesis is that thermal fluctuations of the
spins outside the window at a distance $\xi(t)$ and greater, which are not
equilibrated with respect to spins in the window, effectively generate a
random noise to the spins in the window which is similar to the
random perturbation coming from changing the bonds in the outer region
according to the AW metastate.  It would also be very \textit{useful} if the
metastates were equivalent
because then a theory of the \textit{statics} of spin glasses would give results
corresponding to experiments, which are a \textit{time average}, as is the
case for
simpler systems with a phase transition like ferromagnets.

For the three-dimensional spin glass,
Refs.~\cite{banosetal:10} and \cite{banosetal:10b} have shown that 
an equilibrium calculation in the zero spin-overlap sector gives a power-law decay for
the spin correlations, as
in Eq.~\eqref{alpha_s}, with a value of
$\alpha_s$ consistent with that obtained from dynamics following a quench 
in Ref.~\cite{bellettietal:09}. These are both numerical results.
Here we wish to consider the mean field limit, $d > 6$, because there is an
exact
\textit{analytic} result
$\alpha_s = d - 4$, in RSB theory~\cite{read:14,marinari:00a} to compare with.
Unfortunately it is difficult to carry out useful Monte Carlo simulations
for a spin glass below $T_c$ in more than six dimensions, because the number of sites
in a region of linear size $\ell$, increases so fast, as $\ell^d$, that the
range of $\ell$ that can be studied is very limited. The only calculation of
the exponent $\alpha_d$ in the mean field region that we are aware of is that
of Ref.~\cite{parisi:97} who studied $d = 6$ but only \textit{at} $T=T_c$ and
so these results are for the critical point rather than the spin glass phase.

However, it has been pointed out that a class of models
in one-dimension, with long-range interactions falling off with a power of
distance can serve as a proxy for short-range models
~\cite{kotliar:83,katzgraber:03,banos:12} in a range of dimensions
including high dimension. The Hamiltonian is
\begin{equation}
\mathcal{H} = - \sum_{i, j} J_{ij} S_i S_j \, , 
\end{equation}
where the sites $i = 1, 2, \cdots, N$ lie on a one-dimensional chain with periodic boundary
conditions, the Ising spins $S_i$ take values $\pm 1$, and the interactions
$J_{ij}$ are independent random variables with whose distribution has mean and
variance given by
\begin{equation}
\left[ J_{ij} \right] = 0, \qquad \left[ J_{ij}^2 \right] \propto {1 /
R_{ij}^{2\sigma}} \, ,
\label{Jij}
\end{equation}
in which $\sigma$ is a parameter which can be varied. To incorporate periodic
boundary conditions it is convenient to put the sites on a ring and define
$R_{ij}$ to be the chord distance between $i$ and $j$,
i.e.\ $R_{ij} = (N/\pi)\, \sin(\pi |i-j| / N)$, whereas the distance \textit{along
the ring} is
$r_{ij} = |i-j|$ if $|i-j|<N/2$ and $r_{ij}=N-|i-j|$ otherwise.
The bonds are generated, and the
constant of proportionality in Eq.~\eqref{Jij} fixed, in the following way due 
to Ref.~\cite{leuzzi:08}. We
choose an average coordination number $z_b$, which we take here to be $z_b =
6$. We choose a site $i$ at random and then a site $j$ with a probability
$C/R_{ij}^{2\sigma}$, where $C$ is the normalization constant. If there is
already an interaction between $i$ and $j$ repeat until a pair $(i,j)$
is found which has
not occurred before. Then assign an interaction between $i$ and $j$
chosen from a Gaussian distribution with mean zero and standard deviation
unity. Repeat this $N z_b / 2$ times, so there are $N z_b/2$ interactions. 

Varying $\sigma$ is argued to be analogous to changing the dimension $d$ of a
short-range model~\cite{katzgraber:03}.
In the mean field regime (for the short-range case, $d > d_u = 6$), a precise
connection can be given between $\sigma$ and an equivalent $d$,
namely~\cite{leuzzi:08,katzgraber:09,banos:12},
\begin{equation}
d = 2 / (2 \sigma - 1) \, ,
\label{d_sigma}
\end{equation}
and so, for the long-range model, the mean field regime is $1/2 < \sigma < 2/3$.

\begin{figure}[tb!]
\begin{center}
\includegraphics[width=\columnwidth]{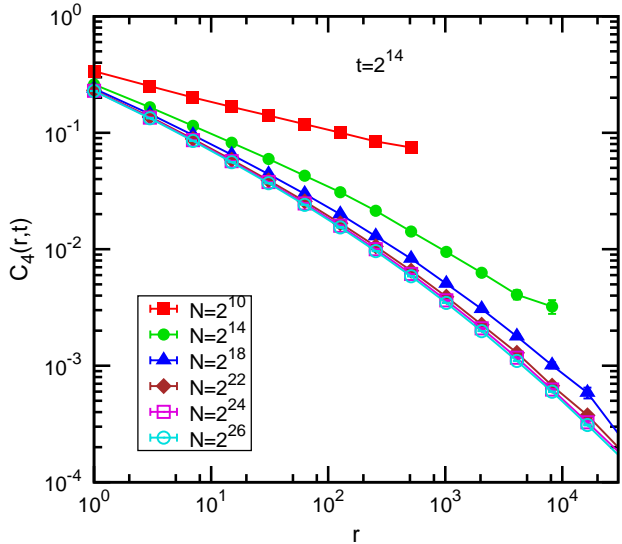}
\caption{
Data for the correlation function $C_4(r, t)$, defined in
Eq.~\eqref{C4rt}, as a function of $r \equiv |i-j|$
for a range of sizes
between $N = 2^{10}$ and $2^{26}$. The data is averaged over about
1000 bond
configurations.
It is also averaged over
times between $2^{14}$ and half that value. There are clearly strong finite-size
effects but the data seems to have converged for the largest sizes, at least
up to distances of order $10^4$.
\label{Fig:t14}
}
\end{center}
\end{figure}

The connection between critical exponents of the short-range and corresponding
long-range models has been discussed systematically in Ref.~\cite{banos:12},
where it is noted that
an exponent of the short-range model in $d$ dimensions
is $d$ times the corresponding exponent of the equivalent
one-dimensional long-range model. Hence, to get the exponent $\alpha_s = d
- 4$ in
the static metastate for the long-range model we divide by $d$ and, since we
work
in the mean field regime, use Eq.~\eqref{d_sigma} to relate $d$ to $\sigma$. This
gives
\begin{equation} 
\alpha_s = 3 - 4 \sigma \quad \text{(long-range model)} .
\label{alphas_LR}
\end{equation}

Here we focus on one value in the mean-field regime, $\sigma = 5/8$,
which corresponds to $d = 8$ according to Eq.~\eqref{d_sigma}.
Using standard finite-size scaling analysis we
find that $T_c  = 1.85(2)$ for this model with $z_b = 6$. 
We need to work \textit{well} below $T_c$ so that our data is characteristic
of the ordered phase and does not also incorporate critical fluctuations.  We
take $T = 0.4 T_c = 0.74$.  We have preferred to focus the numerical effort,
which is substantial, on one temperature in order to get the best quality
data for the largest possible range of sizes. 

We quench the system from infinite temperature to $T = 0.74$ at time $t = 0$
and follow the evolution of the system using Monte Carlo simulations. We
measure spin correlations, averaging them for times between $2^k$ and $2^{k-1}$,
for integer $k$ up to a maximum value.
For the largest sizes this was
$k = 14$.
We find that finite-size effects
are very large and we need to study enormously large sizes. We therefore
take a range of sizes which also increases geometrically, $N = 2^\ell$ up to
$\ell = 26$. We also average over about 1000 samples (the precise number
depending on size).

Figure \ref{Fig:t14} shows our data for for the correlation function $C_4(r,
t)$, defined in Eq.~\eqref{C4rt}, as a function of $r \equiv |i-j|$ at $t =
2^{14}$ for different sizes. Despite the strong finite-size effects the data
seems to have converged for the largest sizes at least for the range of
distance presented. 

\begin{figure}[tb!]
\begin{center}
\includegraphics[width=\columnwidth]{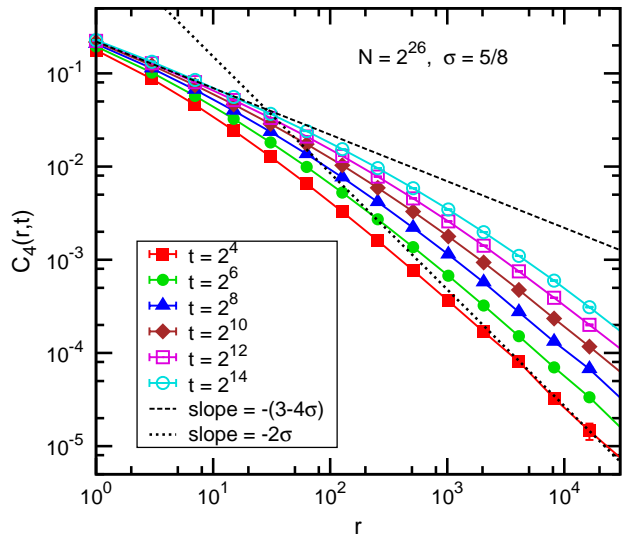}
\caption{
Data for the correlation function for the largest size $N = 2^{26}$ as a
function of $r$ for different times between $t = 2^4$ and $2^{14}$.
A gradual crossover can be seen between two power laws. At large $t$ and small $r$,
$C_4(r, t) \propto 1/r^{\alpha_d}$ with $\alpha_d = 3 - 4 \sigma$ (which
is also the value of $\alpha_s$ expected from the static metastate using
RSB theory). This is the region in which the data has reached a steady
state. At large $r$ and small $t$ 
one finds $C_4(r, t) \propto 1/r^{2\sigma}$ which is just the 
average of the square of the interactions $J_{ij}$.
\label{Fig:N26}
}
\end{center}
\end{figure}

Having established that the largest size, $N = 2^{26}$, is large enough to
eliminate finite-size effects for the range of $r$ and $t$ being considered we
now discuss the data for this size in more detail. Figure \ref{Fig:N26} shows
data for $C_4(r, t)$ at different times as a function of $r$. It is expected to
have the scaling form shown in Eq.~\eqref{C4rt}. For short range models the
scaling function $f(x)$ decays exponentially at large $x$ because the
correlation function falls off very rapidly once $r$ is greater than the
dynamic correlation length. However, in the present model we have interactions
of arbitrarily long range which give a ``direct'' contribution to the
correlation function at large distances. Since $C_4(r, t)$ involves the square
of the spin-spin correlation function, and is averaged over the interactions,
the direct contribution should be proportional to
$[J^2_{ij}]$, which, according to Eq.~\eqref{Jij}, is  proportional
to $\propto r^{-2\sigma} \ (= r^{-5/4}$ for $\sigma = 5/8$).
Fig.~\ref{Fig:N26} follows this behavior at short times and large distances, see
the dotted line.

\begin{figure}[tb!]
\begin{center}
\includegraphics[width=\columnwidth]{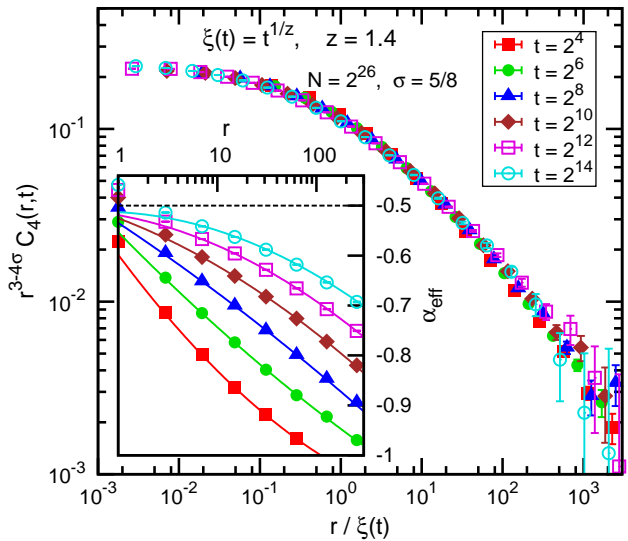}
\caption{
The main part of the figure is a
scaling plot of our data for the largest size $N = 2^{26}$ at $T= 0.74$ according to
Eqs.~\eqref{C4rt} and \eqref{xizT}. We assume a pure power law for $\xi(t)$,
namely with $\xi(t) = t^{1/z}$.
The data collapses well with a dynamic
exponent $z(T) = 1.4$. The inset is a plot of the effective exponent
$\alpha_\text{eff}$, the slope of the curves in Fig.~\ref{Fig:N26}, obtained by
differentiating a spline fit. The lines are quadratic fits to the data for
intermediate $r$ ($7 \le r \le 255$). One sees corrections to the parabolic
fits at very small distances, $r \le 3$.
\label{Fig:C4scale}
}
\end{center}
\end{figure}

By contrast, at small $r$ and large $t$, where $r \ll \xi(t)$,
the data  for different times collapses and is
consistent with a
decay
$r^{-(3 - 4 \sigma)}\ (= r^{-1/2}$ for $\sigma = 5/8$),
see the dashed line in Fig.~\ref{Fig:N26}. To estimate better the slope at
large $t$ and small $r$ we plot in the inset to Fig.~\ref{Fig:C4scale}, the effective
exponent $\alpha_\text{eff}$, the slope of the data in
Fig.~\ref{Fig:N26}, as a function of $r$ for different times. The curves are
quadratic fits for intermediate  $r$ ($7 \le r \le 255$). The intercepts of
the fits approach $-0.5$ for $r \to 0$ at large $t$.
Hence, according
to Eq.~\eqref{alpha_d}, we have
$\alpha_d = 3 - 4 \sigma$ (or at least very close to it.)
However, this is precisely equal to $\alpha_s$, the corresponding exponent from the
\textit{static}
metastate according to RSB theory as shown in Eq.~\eqref{alphas_LR}. Hence we
see that, in the mean field regime, the static and dynamic metastates
appear to agree and the description appears to be that of RSB.
The latter agrees with some other studies~\cite{katzgraber:05},
and is also implied by those, such as Refs.~\cite{banosetal:10,banosetal:10b},
which argue that RSB
holds even below six dimensions.

The main part of
Fig.~\ref{Fig:C4scale} shows a scaling plot of our data for the largest size
according to Eqs.~\eqref{C4rt} and \eqref{xizT}. The data scales well and
indicates
$z(0.4 T_c) = 1.4(2)$.
For short-range models $z(T)$ is found to obey
Eq.~\eqref{zT}.
If we assume the same here then $z(T_c) = z_c = 0.56 (8)$.
To translate this value for $z_c$, the \textit{critical} dynamical exponent,
into the exponent for the equivalent
short-range model, we multiply by $d \ (=8)$, as discussed 
above, so our
estimate for the critical dynamical exponent of the $d = 8$ short-range spin
glass is $z_c = 4.5(6)\ (d = 8)$.
This model is in the mean field region ($d > 6$) for which the dynamical
exponent is found to be $z_c = 4$~\cite{zippelius:84,parisi:97}. Our result 
is consistent with this.

To conclude, we have shown quantitatively that the non-equilibrium dynamics
following a quench of a model which is a proxy for a short-range spin
glass in dimension $d > 6$ is given, in the steady-state regime
where the distance is less
than the non-equilibrium correlation length, by the
\textit{analytic} result for the \textit{static} metastate calculated
according to RSB theory.
This suggests that
(i)
RSB theory applies to spin glasses above the upper critical dimension, $d_u = 6$, and
(ii)
the dynamic and static metastates are equivalent (at least in this region). If
the latter is true \textit{in general}
it would provide a great simplification in the
study of spin glasses.


\begin{acknowledgments}
This work is supported in part by the National
Science Foundation under Grant
No.~DMR-1207036 and by
a Gutzwiller Fellowship for one of us (APY) at the Max Planck Institute
for the Physics of Complex Systems (MPIPKS), Dresden. We also thank
the MPIPKS for a generous allocation of computer time which made possible
these simulations. We thank Nick Read for an illuminative exchange on the
subject of the metastate, and
also thank him, Victor Martin-Mayor, Alexander Hartmann, Helmut
Katzgraber, and Daniel Fisher
for helpful comments on an
earlier version of the manuscript.
\end{acknowledgments}

\bibliography{refs,comments}

\end{document}